\documentclass[10pt]{article}
\usepackage{psfig,epsf,conf-X}

\begin{document} 
\small
\heading{%
%
Testing for substructure in optical and X--ray clusters
}
\par\medskip\noindent
\author{%
V. Kolokotronis$^{1}$, S. Basilakos$^{1,2,3}$, M. Plionis$^1$, I. 
Georgantopoulos$^1$, 
}
\address{%
Astronomical Institute, National Observatory of Athens, I Metaxa
$\&$ B. Pavlou, Palaia Penteli, 
15236, Athens, Greece
}
\address{%
Physics Department University of Athens, Panepistimiopolis, Zografos,
Athens, Greece 
}
\address{%
Imperial College of Science, Technology and Medicine, Blackett Laboratory, 
Prince Consort Road, London SW1 2EZ, UK
}

\begin{abstract}
We present a detailed study of the morphological features of 22 rich galaxy 
clusters. We systematically compare cluster images and morphological 
parameters in an attempt to reliably identify possible substructure in both 
optical and X--ray images. To this end, we compute moments of the 
surface-brightness distribution to estimate ellipticities, center-of-mass 
shifts and orientations.
We find important correlations between the optical 
and X--ray morphological shape parameters. Most of our clusters (17) have a 
good 1-to-1 correspondence between the optical and the X--ray images and at 
least 9 appear to have strong indications of substructure. This corresponds 
to a percentage of $\sim 40 \%$ in good accordance with other 
similar analyses. Finally, 4 out of 22 systems seem to have distinct 
subclumps in the optical which are not verified in the X--ray images, and 
thus are suspect of being due to optical projection effects.
We assess the significance of results using Monte Carlo simulations. 

\end{abstract}
\section{Introduction}

One of the most significant properties of galaxy clusters is the relation 
between their dynamical state and the underlying cosmology. In an open 
universe, clustering effectively freezes at high redshifts and clusters today 
should appear more relaxed with weak or no indications of substructure. 
Instead, in a critical density model, such systems continue to form even 
today and are expected to be dynamically active. The percentage and 
morphologies of disordered objects in a cluster sample could lead to crucial 
constraints on $\Omega_{\circ}$ and $\Lambda$, especially if combined with 
N-body/gas-dynamic numerical simulations spanning different dark matter (DM) 
scenarios (\cite{RLT92}; \cite{Evr}; \cite{Thomas}).

A large number of relevant analyses have been devoted to this study and an 
accordingly varying and large number of optical and X--ray cluster 
compilations have been utilised to this aim (\cite{M95}; \cite{BT96}; 
\cite{JF99} and references therein). In the present work, we use a sample of 
22 galaxy clusters (APM and ROSAT) in a complementary fashion with the aim to 
address the following two questions: 

\begin{itemize}
\item Is substructure in the X--ray also corroborated by the optical 
observations and in what percentage?
\item  What is the percentage of systems depicting strong indications of 
subclumping and what does it imply for the existing cosmology?
\end{itemize}

\section{Data \& Methodology}
The present dataset follows from a double cross-correlation between rich 
ACO clusters ($\rm R\geq\;$1,2,3) with the APM cluster catalogue and the 
X--ray (0.1 - 2.4) keV ROSAT pointed observations archive, finally resulting 
in 27 common entries. Due to problematic regions of the APM catalogue, low 
signal to noise X--ray observations and contamination by known foreground or 
background objects, we exclude 5 clusters reducing our cluster sample to 22 
systems. The redshift range of our sample is $0.04 \leq z \leq 0.13$ with 
$\langle z \rangle \sim 0.074$ and median $\sim 0.069$. For the needs of our 
analysis we transform cluster redshifts to distances using the 
luminosity-distance relation for a critical density model, $q_{\circ}=0.5$ 
and $H_{\circ}=100 \;h$ km $s^{-1}$ Mpc$^{-1}$.
\begin{figure}[h]
\hskip -2cm
\mbox{\epsfysize=10cm \epsffile{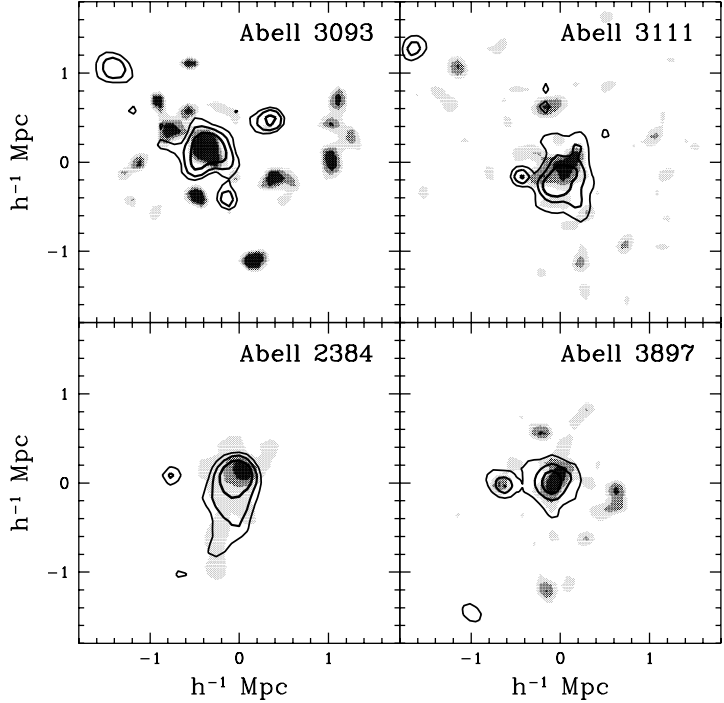}}
\caption[]{ Optical and X--ray images of galaxy clusters. Contours correspond 
to the X--ray data, whereas greyscale configurations denote the optical.}
\end{figure}

In order to construct a common comparison base, we create a continuous 
density field for both optical and X--ray data by using a Gaussian Kernel and 
a variable smoothing length according to each cluster redshift. However, so 
as to take into account the reduction of the number of cluster members as a 
function of distance (due to the APM magnitude limit), and thus the 
corresponding increase of discreteness effects, we have investigated, using 
Monte-Carlo cluster simulations, the necessary size of the smoothing window 
in order to minimise such effects and optimize the performance of our 
procedure (cf. \cite{BPM}). We then compute the optical and X--ray cluster 
shape parameters utilising the method of moments of inertia. The eigenvalues 
and the eigenvectors of the inertia tensor can provide us with the cluster 
ellipticities and major axis orientations (position angles) respectively. We 
also define the centroid shift as the vectorial difference between the 
weighted cluster center-of-mass and the highest cluster density peak (cf. [3],
[4]). These shape parameters are estimated using all cells that have 
densities above three thresholds. These are defined as the average density of 
all cells that fall within a chosen radius. The three radii used are 
$r_{\rm \rho}=0.3, 0.45$ and $0.6 \;h^{-1}$ Mpc, whereas the maximum 
searching radius for all subsequent calculations is $0.8h^{-1}\;$Mpc. We 
finally utilise a friend-of-friends algorithm to investigate possible 
substructure by joining all cells having common boundaries and fall above 
each density threshold. We therefore create and register all subgroups as a 
function of density threshold and rank substructure according to different 
criteria (cf. \cite{BPM}). 

\section{Quantifying substructure results}

Looking at the cluster shape parameters and visually inspecting the isodensity 
contour maps (see Figure 1 for a subsample of 4 objects), we do observe a 
remarkable 1-to-1 correspondence in $\leq 80\%$ of our sample regarding the 
gross structural features (prime and secondary components, elongations, 
irregular activity, collision vestiges, unimodality). The majority of the 
optical and X--ray images are very well aligned with 
$\langle \delta \theta \rangle \leq 20^{\circ}$ and relative correlation 
coefficient of order of $\ge 0.9$, 
which is also highly significant. Furthermore, 
the ellipticities and the centroid shifts between optical and X--ray data do 
correlate well with coefficient of order $\sim 0.7$ in both cases. On the 
other hand, we have found important intrinsic correlations between 
ellipticities and centroid variations in the optical and the X--ray 
configurations separately, with relative coefficients ranging from 0.6 to 0.8 
respectively. Cross-correlating the optical and X--ray substructure measures, 
we discover that they also correlate nicely. 
Probably the most interesting correlation is that between optical centroid
shifts and X--ray ellipticities, with a value exceeding 0.8. This indicates
that we can deduce 
the shape of the DM gravitational potential from optical cluster data.  

Since random density fluctuations as well as background contamination
may introduce spurious substructure, we quantify the significance of our 
substructure measures, as revealed by the center of mass shift in the optical,
using Monte Carlo cluster simulations with the same number
of galaxies, ellipticity and estimated background as that of each cluster in
our sample. This significance is estimated 
by measuring the deviation of the true cluster center-of-mass shift, from the 
corresponding simulated 
value in units of the estimated $\sigma$ from 100 Monte-Carlo simulations of
each cluster.
We also compare this significance measure to the results of the subgroup 
statistics algorithm. The two measures are significantly correlated
with a value of $> 0.7$. 

Finally we classify our clusters according to their morphological parameters 
using a scheme which is very close to the one developed by \cite{JF99}. 
Results indicate that our findings are in very good agreement with those of 
\cite{JF99} both on a quantitative and qualitative basis. Note also, that 11 
out of our 22 clusters have been examined for substructure signals elsewhere 
in the literature. We have checked that our computations on the cluster shape 
parameters do accord with those of the other studies. (\cite{M95}; 
\cite{BT96}; \cite{JF99}) 

From our prime substructure analysis we confirm that at least 9 out of 22 
systems display strong substructure indications visible in both parts of the 
spectrum. We also find that 4 clusters ($\leq 20\%$) show clear disparities 
between the optical and X--ray maps, with apparent substructure in the 
optical not corroborated by the X--ray data. The rest of our sample exhibits 
no or insignificant substructure indications. We finally observe that our 
present study is compatible with that of \cite{RLT92} (their Figure 2) 
regarding the 
cluster substructure frequency, setting a rather frail lower limit on the 
density parameter ($\Omega_{\circ}\,\geq\,0.5$).

In the near future we plan to apply the methodology of this work to the large 
optical APM sample of galaxy clusters ($> 900$ entries), in order to 
investigate in more detail the issue of cluster substructure. 

\acknowledgements{V. Kolokotronis and S. Basilakos wish to acknowledge 
financial support from the Greek State Fellowship Foundation.}

\vfill
\end{document}